\documentclass[twocolumn,showpacs,preprintnumbers,amsmath,amssymb]{revtex4}

\usepackage{graphicx}
\usepackage{dcolumn}
\usepackage{bm}

\newtheorem{corollary}{Corollary}
\newtheorem{theorem}{Theorem}
\newtheorem{definition}{Definition}

\newcommand{\bra}[1]{\langle #1|}
\newcommand{\ket}[1]{|#1\rangle}

\begin{document}


\title{Separable Multipartite Mixed States - Operational Asymptotically Necessary and Sufficient Conditions}

\author{Fernando G. S. L. Brand\~ao}
\email{fgslb@ufmg.br}
\author{Reinaldo O. Vianna}
\email{reinaldo@fisica.ufmg.br}
\affiliation{Universidade Federal de Minas Gerais - Departamento  de F\'{\i}sica\\
Caixa Postal 702 - Belo Horizonte - MG -  Brazil - 30.123-970}
\date{\today}


\begin{abstract}
We introduce an operational procedure to determine, with arbitrary probability
 and accuracy, optimal entanglement witnesses for every multipartite entangled
 state. This method provides an operational criterion for separability
 which is asymptotically necessary and sufficient. Our results are also 
generalized to detect all different types of multipartite entanglement.

\end{abstract}

\maketitle

\section*{}

 Entanglement, first noticed by Einstein, Podolsky, and Rosen [1], is at the heart of quantum mechanics. Quantum teleportation, superdense coding and cryptography   [2] are achieved only when one deals with inseparable states. Thus, the determination and quantification of entanglement in a composite quantum state is one of the most important tasks of quantum information theory. A finite-dimensional density operator $\rho_{1...n} \in {\cal B}(H_{1} \otimes ... \otimes H_{n})$ (the Hilbert space of bounded operators acting on $H_{1}\otimes ... \otimes H_{n}$) is separable iff it can be written as a convex sum of separable pure states: 
\begin{equation}
\rho_{1...n} = \sum_{i}p_{i} \ket{\psi_{i}}_{11}\bra{\psi_{i}}\otimes ... \otimes \ket{\psi_{i}}_{nn}\bra{\psi_{i}}
\end{equation}
where ${\cal f}p_{i}{\cal g}$ is a probability distribution and $\ket{\psi_{i}}_{k}$ are vectors belonging to Hilbert spaces $H_{k}$. Despite the simplicity of this definition, no operational necessary and sufficient criterion have been found for the separability problem until now.
 Moreover, it was shown by Gurvits [3] that this problem is NP-HARD. In this letter,  we present a procedure to determine,  with a chosen probability, if a given state is entangled. In order to do that, we apply a class of convex optimization problems known as robust semidefinite programs (RSDP) to the concept of entanglement witness (EW) which we briefly recall.

An operator $\rho_{1...n}$ is entangled iff there exists a self-adjoint operator $W \in {\cal B}(H_{1} \otimes ... \otimes H_{n})$ which detects its entanglement [4], i.e., such that $Tr(W \rho_{1...n}) < 0$ and $Tr(W\sigma_{1...n}) \geq 0$ for all $\sigma_{AB}$ separable. This condition follows from the fact that the set of separable states is convex and closed in ${\cal B}(H_{1} \otimes ... \otimes H_{n})$. Therefore, as a conclusion of the Hahn-Banach theorem, for all entangled states there is a linear functional which separates them from this set. We will deal in this paper only with normalized entanglement witnesses such that $tr(W) = 1$.
\begin{definition}
A hermitian operator $W_{opt} \in {\cal B}(H_{1} \otimes ... \otimes H_{n})$ is an optimal EW for the density operator $\rho_{1...n}$ if
\begin{equation}
Tr(W_{opt}\rho_{1...n}) \leq Tr(W\rho_{1...n})
\end{equation}
for every EW $W$.
\end{definition}
Although this definition of OEW is different from the one introduced in [5], the optimal EWs of both criteria  are equal.

We may now express the search of an optimal EW for an arbitrary state $\rho_{1...n}$ in terms of a robust semidefinite program (RSDP) . A semidefinite program (SDP) consists of minimizing a linear objective under a linear matrix inequality (LMI) constraint, precisely, 
\begin{center}
minimize $c^{\cal y}\textbf{x}$ subject to
\end{center}
\begin{equation}
F(\textbf{x}) = F_{0} + \sum_{i=1}^{m}x_{i}F_{i} \geq 0
\end{equation}
where $c \in {\cal C}^{m}$ and the hermitian matrices $F_{i} = F_{i}^{\cal y} \in {\cal C}^{n x n}$ are given and $x \in {\cal C}^{m}$ is the vector of optimization variables. $F(x) \geq 0$ means $F(x)$ is hermitian and positive semidefinite. SDPs are global convex optimization programs and can be solved in polynomial time with interior-point algorithms [6]. For instance, if there are $m$ optimization variables and $F(x)$ is a n\hspace{0.03 cm}x\hspace{0.03 cm}n matrix, the number of operations scales with problem size as $O(m^{2}n^{2})$. SDPs have already been used in different problems of quantum information theory [7] and also in the separability problem [8]. An important generalization of $(3)$ is when the data matrices $F_{i}$ are not constant, i.e., they depend of a parameter which varies within a certain subspace. This family of problems, known as robust semidefinite programs, is given by:
\begin{center}
minimize $c^{\cal y}\textbf{x}$ subject to
\end{center}
\begin{equation}
F(\textbf{x}, \Delta) = F_{0}(\Delta) + \sum_{i=1}^{m}x_{i}F_{i}(\Delta) \geq 0, \hspace{0.2cm} \forall \Delta \in {\cal D}
\end{equation}
where ${\cal D}$ is a given vectorial (sub)space. Note that problem $(4)$ is more difficult to solve than $(3)$, since one must find an optimization vector $x$ such that $F(x, \Delta)$ is positive semidefinite for all $\Delta \in {\cal D}$. One often encounters SDPs in which the variables are matrices and in which the inequality depends affinely on those matrices. These problems can be readily put in the form $(3)$ by introducing a base of hermitian matrices for each matrix variable. However, since most of optimization solvers [9] admit declaration of problems in this most general form, it is not necessary to write out the LMI explicitly as $(3)$, but instead make clear which matrices are variables. Equality constraints involving the optimization variables can also appear in $(3)$ and $(4)$ without any further computational effort. We can now enunciate the main result of this letter.
\begin{theorem}
A state $\rho_{1...n} \in {\cal B}(H_{1} \otimes ... \otimes H_{n})$ is entangled, i.e., can not be decomposed as $(1)$, iff the optimal value of the following RSDP is negative:
\begin{center}
minimize $Tr(W\rho_{1...n})$ subject to
\end{center}
\begin{equation}
\sum_{i_{1}=1}^{d_{n}}\sum_{j_{1}=1}^{d_{n}} ... \sum_{i_{n-1}=1}^{d_{n}}\sum_{j_{n-1}=1}^{d_{n}} \left ( a_{i_{1}}^{*}...\hspace{0.05cm} a_{i_{n-1}}^{*} a_{j_{1}}... \hspace{0.05cm}a_{j_{n-1}} \right.
\end{equation}
\[ \left. W_{i_{1}... \hspace{0.01 cm}i_{n-1}j_{1}... \hspace{0.01 cm}j_{n-1}} \right) \geq 0 \]
\[ Tr(W) = 1, \hspace{0.2cm} \forall a_{i_{k}} \in {\cal C},\hspace{0.2cm} 1 \leq k \leq n - 1\]
where $d_{n}$ is the dimension of $H_{n}$, $W_{i_{1}...\hspace{0.01cm}i_{n-1}j_{1}...\hspace{0.01cm}j_{n-1}} =  {}_{1}\bra{i} \otimes ... \otimes {}_{n-1}\bra{i}W\ket{j}_{n-1} \otimes ... \otimes \ket{j}_{1} \in {\cal B}(H_{1} \otimes ... \otimes H_{n-1})$ and $\ket{j}_{k}$ is an orthonormal base of ${\cal H}_{k}$. If $\rho_{1...n}$ is entangled, the solution matrix $W$ which minimizes $Tr(W\rho_{1...n})$ is the OEW for $\rho_{1...n}$.
\end{theorem}
{\bfseries proof}: First we have to show that $(5)$ is a genuine RSDP. Note that $W_{i_{1}... \hspace{0.01cm}i_{n-1}j_{1}... \hspace{0.01 cm}j_{n-1}}$ and the objective $Tr(W\rho_{1...n})$ are both linear in the matrix variable $W$. Thus $(5)$ can be put in the form $(4)$, where ${\cal D}$, in this case, is ${\cal C}^{d_{n}}$. A state $\rho_{1...n}$ is entangled iff there exists an operator $W$ such that $Tr(W\rho_{1...n}) \leq 0$ and $_{1}\bra{\psi}\otimes ... \otimes _{n}\bra{\psi}W\ket{\psi}_{n}\otimes ... \otimes \ket{\psi}_{1}\geq 0$ for all states $\ket{\psi}_{k} \in H_{K}$. Therefore, the matrix $_{1}\bra{\psi}\otimes ... \otimes _{n-1}\bra{\psi}W\ket{\psi}_{n-1}\otimes ... \otimes \ket{\psi}_{1}\geq 0$ has to be semidefinite positive for all $\ket{\psi}_{k} \in H_{K}$ . Letting $\ket{
\psi}_{k} = \sum_{j}a^{k}_{j}\ket{j}_{k}$, where $\ket{j}_{k}$ is an orthonormal base of ${\cal H}_{k}$, it is straightforward to show that the optimal W given by $(5)$ is the OEW of $\rho_{1...n}$. {\bfseries QED}.

In spite of the similarity between $(3)$ and $(4)$, RSDPs are in general very hard optimization problems. Actually, it was proved that robust semidefinite programs in the form of $(5)$ are NP-HARD [10]. 
\begin{corollary}
The determination of the OEW for an arbitrary state $\rho_{1...n}$ is a NP-HARD problem.
\end{corollary}
Since $(5)$ is computationally intractable, it is natural to search for approximations of it in terms of SDPs, which are very efficiently solved. These relaxations of RSDP have been intensively studied [11] in the past years and can be classified as deterministic or probabilistic. In this letter we will focus on the latter, where one seeks a feasible solution to most of the possible values of the varying parameters. The results of applying deterministic relaxations to $(4)$, which yields  new separability sufficient criteria, was reported in [12]. Our methodology will be based on the concept of $\epsilon$-level solution introduced in [13]. 

Consider the most general form of RSDP given by $(4)$. Assume that the support ${\cal D}$ for $\Delta$ is endowed with a $\sigma$-algebra and that a probability measure $\textbf{P}$ over this algebra is also assigned. Let $x \in {\cal C}^{m}$ be a candidate solution to $(4)$. The probability of violation of x is defined as: $V(x) = \textbf{P}{\cal f}\Delta \in {\cal D}: F(x, \Delta) \leq 0 {\cal g}$. For example, in $(5)$, where the varying parameters are uniformly distributed over ${\cal C}^{d_{n}}$, $V(x)$ measures the percentage of parameters such that the linear matrix inequality is violated. 
\begin{definition} 
Let $\epsilon \in [0,1]$. We say that a hermitian operator W is an $\epsilon$-level entanglement witness, $\epsilon$-W, if
\[ V(W) = \textbf{P}{\cal f}\sigma \in {\cal S}: Tr(W\sigma) < 0 {\cal g} \leq \epsilon \]
where ${\cal S}$ is the subspace of separable density operators. 
\end{definition}
The concept of optimal $\epsilon$-level entanglement witness is totally analogous to the one of definition $(1)$, but now $(2)$ has to hold for every
$\epsilon$-level EW. The importance of this new class of hermitian operators is that, in contrast to the case of genuine EW, $\epsilon$-level optimal EW can be determined with a priori chosen probability in polynomial time for every multipartite state.
\begin{theorem}
Let $\epsilon \in [0,1]$, $\beta \in [0,1]$ and $N \geq \frac{D(D + 1)}{\epsilon \beta} - 1$, where  $D$ is the dimension of $H_{1} \otimes ... \otimes H_{n}$. Assume that N independent identically uniformly distributed samples $a_{j_{l}}^{1}, a_{j_{l}}^{2}, ..., a_{j_{l}}^{N}$, $1 \leq j_{l} \leq d_{n}$ and $1 \leq l \leq n - 1$, are drawn. Then the optimal $\epsilon$-EW for a state $\rho_{1...n}$ is given with probability at least $1 - \beta$ by the solution of the following semidefinite program:
\begin{center}
minimize $Tr(W\rho_{1...n})$ subject to
\end{center}
\begin{equation}
\sum_{i_{1}=1}^{d_{n}}\sum_{j_{1}=1}^{d_{n}} ... \sum_{i_{n-1}=1}^{d_{n}}\sum_{j_{n-1}=1}^{d_{n}} \left ( (a_{i_{1}}^{k})^{*}...\hspace{0.05cm} (a_{i_{n-1}}^{k})^{*} a_{j_{1}}^{k}... \hspace{0.05cm}a_{j_{n-1}}^{k} \right.
\end{equation}
\[ \left. W_{i_{1}... \hspace{0.01 cm}i_{n-1}j_{1}... \hspace{0.01 cm}j_{n-1}} \right ) \geq 0 \]
\[ Tr(W) = 1, \hspace{0.4 cm} 1 \leq k \leq N \]
where $d_{n}$ is the dimension of $H_{n}$, $W_{i_{1}...\hspace{0.01 cm}i_{n-1}j_{1}...\hspace{0.01 cm}j_{n-1}} =  {}_{1}\bra{i} \otimes ... \otimes {}_{n-1}\bra{i}W\ket{j}_{n-1} \otimes ... \otimes \ket{j}_{1} \in {\cal B}(H_{1} \otimes ... \otimes H_{n-1})$ and $\ket{j}_{k}$ is an orthonormal base of ${\cal H}_{k}$.
\end{theorem}
{\bfseries proof}: According to [14], an $\epsilon$-level solution of a RSDP can be obtained with probability $1 - \beta$ from a sampled convex program, where the robust linear matrix inequality is replaced by $N \geq \frac{r}{\epsilon \beta} - 1$ independent identically distributed samples chosen according to probability \textbf{P}, where r is the number of optimization variables of the problem. The result follows in a straightforward manner if one notices that problem $(5)$ has $D(D + 1)$ optimization variables (the number of distinct real entries of W) and that \textbf{P} in this case is uniform. {\bfseries QED}.

Notice that theorem $(2)$ gives a sufficient condition for separability, which is asymptotically  also necessary. In fact, it is possible to determine, with any desired precision and probability, if any state is entangled or not. Nevertheless, one must always consider the trade-off between the accuracy of the results and computation effort. Although a priori feasibility levels are given by the former theorem, the optimization problem yields in general  much better results. Once a solution has been determined, it is possible to make an improved estimate of the level of feasibility using Monte-Carlo techniques. In order to do that, generate a new set of $\tilde{N}$  independent identically uniformly distributed samples $a_{j_{l}}^{k}$ and construct the empirical probability of constraint violation, $V_{emp}(W) = \frac{1}{\tilde{N}}\sum_{i=1}^{\tilde{N}}1(_{1}\bra{\psi}\otimes ... \otimes _{n}\bra{\psi}W\ket{\psi}_{n}\otimes ... \otimes \ket{\psi}_{1} < 0$), where 1(.) is the indicator function. Then, the classical Chernoff inequality guarantees that $|V(W) - V_{emp}(W)| \leq  \epsilon$ holds with confidence grater than $1 - \beta$ , provided that 
\begin{equation}
\tilde{N} \geq \frac{log 2/ \beta}{2 \epsilon^{2}}
\end{equation}
samples are drawn. Another important performance parameter is the minimum eigenvalue over the violated constraint. It can also be obtained empirically and it is very useful to determine if the solution obtained is accurate. 

We present now the first example for which we applied our techniques to determine an approximate optimal entanglement witness. We used MATLAB and the package SEDUMI [9] to implement and solve the SDP. Consider the Horodecki 3 X 3 bound entangled states [15]:
{\small
\begin{equation}
\rho(a) = \frac{1}{8a + 1} \left [
\begin{array}{ccccccccc}
a & 0 & 0 & 0 & a & 0 & 0 & 0 & a \\
0 & a & 0 & 0 & 0 & 0 & 0 & 0 & 0 \\
0 & 0 & a & 0 & 0 & 0 & 0 & 0 & 0 \\
0 & 0 & 0 & a & 0 & 0 & 0 & 0 & 0 \\
a & 0 & 0 & 0 & a & 0 & 0 & 0 & a \\
0 & 0 & 0 & 0 & 0 & a & 0 & 0 & 0 \\
0 & 0 & 0 & 0 & 0 & 0 & \frac{1 + a}{2} & 0 & \frac{\sqrt{1 - a^{2}}}{2} \\
0 & 0 & 0 & 0 & 0 & 0 & 0 & a & 0 \\
a & 0 & 0 & 0 & a & 0 & \frac{\sqrt{1 - a^{2}}}{2} & 0 & \frac{1 + a}{2} \\
\end{array}
\right ]
\end{equation}
}
where $a \in [0,1]$. This family of states is particularly interesting because the Peres-Horodecki criterion fails to detect its entanglement, i.e. they are positive partial transpose entangled states. Using $N = 1200$ samples in each test, we were able to detect entanglement for all values of a, except for 0 and 1. The expectation value of the OEW for each $\rho(a)$ is shown in Fig. 1. The empirical probability ($V_{emp}$(W)) of violation and the minimum eigenvalue over the violated constraint ($\lambda_{min}$), calculated using $\tilde{N} = 10^{6}$ samples, were both negligible, showing that the algorithm converged.
\begin{figure}[h]
\includegraphics{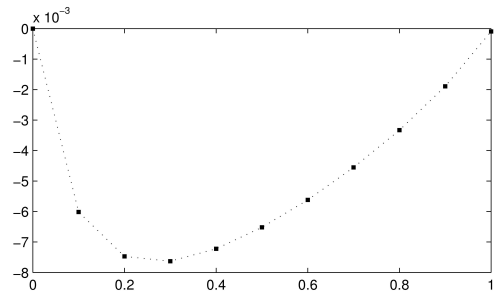}
\caption{  Tr($W_{opt}\rho(a)$) X a, for the 3 X 3 Horodecki bound entangled states.}
\label{fig:PS}
\end{figure}

We have applied our methodology to a large number of 2 x 2 and 2 x 3 states, namely, 5000 (five thousand) random states of each kind. Since in this case the positive partial transpose criterion [16] gives sufficient and necessary conditions for entanglement, we were able to test the reliability of our results. The percentage of misleading conclusions as a function of the number of samples used in the SDP is plotted in Fig. 2. Notice that for $N > 500$ no mistake was made.

\begin{figure}[h]
\includegraphics{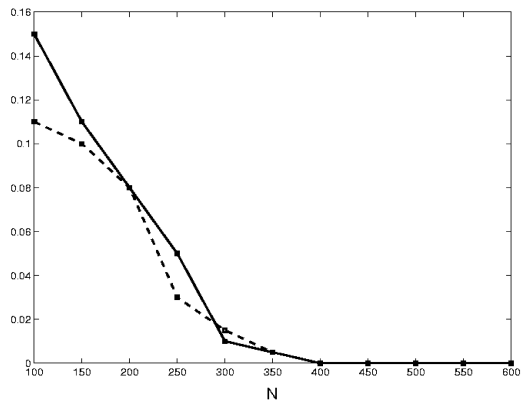}
\caption{  Percentage of wrong results \textbf{X} number of samples (N), for 2x2 (dashed line) and 2x3 (solid line) systems.}
\label{fig:PS}
\end{figure}

As a third example, we will analyze a three-partite bound entangled state derived from the context of the unextendible product bases (UPB) [18]. Consider the complementary state to the \textbf{Shifts} UPB: ${\cal f} \ket{0, 1, +}, \ket{1, +, 0}, \ket{+, 0, 1}, \ket{-, -, -} {\cal g}$, where $\pm = (\ket{0} \pm \ket{1})/\sqrt{2}$. We have calculated the OEW for the three bipartite partitions and for the three-partite partition. The results of the computation with 2000 samples are summarized in Table I.
\begin{table}[h]\centering
\begin{tabular}{|c|c|c|c|} \hline 
\bf{Partition} & \bf{Tr($W_{opt}\rho$)} & \bf{V(W)} & \bf{$\lambda_{min}$} \\ \hline
A-BC & $-3.89x10^{-6}$ & 0,063  & $-4.34x10^{-6}$ \\ \hline
B-AC & $-5.78x10^{-6}$ & 0,040  & $-5.78x10^{-6}$ \\ \hline
C-AB & $-1.12x10^{-6}$ & 0,087  & $-3.69x10^{-6}$ \\ \hline
A-B-C& $-3.17x10^{-3}$ & 0,002  & $-9.23x10^{-7}$ \\ \hline
\end{tabular}
\label{tab:1}
\caption{Results of the method for the three-partite bound entangled state complementary to the Shifts UPB}
\end{table} 
We can conclude that the state is separable with respect to the bipartite splits, whereas it is entangled with respect to tripartite product states.  These same results were obtained using a different approach in [18].

We have considered so far only the discrimination between entangled and separable states. Actually, the structure of multipartite quantum entanglement is much richer [17]. A n-partite density operator $\rho_{1...n}$ is a m-separable state if it is possible to find a decomposition to it such that, in each pure state term, at most m parties are entangled among each other, but not with any member of the other group of $n - m$ parties. Furthermore, even in the class of m-separable states, there exist different types of entanglement, i.e, states which cannot be converted to each other by local operations and classical communication protocols (LOCC). 
Since the subspace of m-separable density operators is convex and closed, it is also possible to apply the Hahn-Banach theorem to it and establish the concept of entanglement witness to $(m+1)$-partite entanglement. In order to do that, consider the index set $P = {\cal f}1, 2, ..., n{\cal g}$. Let $S_{i}$ be a subset of P which has at most m elements. Then W is an $(m+1)$-partite entanglement witness if:
\begin{equation}
\begin{array}{c}
 _{S_{i_{v}}}\bra{\psi}\otimes ... \otimes \hspace{0.07 cm}_{S_{i_{1}}}\bra{\psi}W\ket{\psi}_{S_{i_{1}}} \otimes ... \otimes \ket{\psi}_{S_{i_{v}}} \geq 0 \\
\forall \hspace{0.2 cm} S_{i_{1}}, ..., S_{i_{v}} \hspace{0.2 cm} $such that$ \\
\bigcup_{k=1}^{v}S_{i_{k}} = P \hspace{0.2 cm} $and$ \hspace{0.2 cm} S_{i_{k}} \bigcap S_{i_{l}} = {\cal f}{\cal g} 
\end{array}
\end{equation}
Therefore, it is possible to apply the same methods developed earlier to $(m + 1)$-partite EW, where one has to minimize $Tr(W\rho_{1...n})$ subject to the RSDP derived from (8).

As a final example, we determine a tripartite-entanglement OEW for the GHZ state $\ket{\psi_{GHZ}} = \frac{1}{\sqrt{2}}(\ket{000} + \ket{111})$, i.e., an operator which separates $\ket{\psi_{GHZ}}$ from the set of bi-separable density matrices. Also in this case, using $N = 2000$ samples, our procedure has found a genuine OEW for the state:
{\small
\[ W_{opt} = \frac{1}{6}(\ket{001}\bra{001} + \ket{010}\bra{010} + \ket{011}\bra{011}\] 
\[ + \ket{100}\bra{100} + \ket{101}\bra{101} + \ket{110}\bra{110} - \ket{000}\bra{111} \] \[- \ket{111}\bra{000}) \] 
}
One can easily check that this matrix is indeed positive semidefinite over the separable states. 

In summary, we have constructed a procedure to determine with arbitrary probability and accuracy optimal entanglement witness for every entangled state. Thus, considering the NP-hardness of the separability problem, this approximate method is of great importance to the development of the theory of entanglement. The search of others approximate algorithms for the optimization of EW with improved performance is an interesting problem for further research.

\begin{acknowledgments}

Financial support from the Brazilian agencies CNPq, Institutos do Mil\^enio-Informa\c{c}\~ao Qu\^antica(MCT) and  FAPEMIG.

\end{acknowledgments}

\end{document}